\documentclass[namedreferences,hyperref,optionalrh,solaromanenum]{spr-sola}
\usepackage{siunitx}

\usepackage{graphicx}                    %
\usepackage{color}                       %
\usepackage{amsmath}    %
\usepackage{xcolor}

\usepackage{lineno}
\usepackage{microtype}

\chardef\us=`\_

\DeclareSIUnit\keV{\kilo\electronvolt}
\DeclareSIUnit{\counts}{counts}
\DeclareSIUnit{\channel}{channel}
\DeclareSIUnit\erg{erg}

\usepackage[final]{changes}
\definechangesauthor[name={Abhilash}, color=red]{ARS}

\begin{document}

\begin{frontmatter}

\title{Iron Fluorescence in X-class Solar Flares: Aditya-L1/SoLEXS Observations}

\author[addressref={ursc,iitg},corref,email={sarwade@ursc.gov.in}]{\snm{Abhilash~R.~Sarwade}\orcid{0009-0002-0216-0545}}
\author[addressref={ursc,cessi},email={sankark@ursc.gov.in}]{\snm{K.~Sankarasubramanian}\orcid{0000-0003-1406-4200}}
\author[addressref=ursc]{\fnm{Monoj}~\snm{Bug}}
\author[addressref=ursc]{\fnm{Vaishali}~\snm{Sharan}}
\author[addressref=ursc]{\fnm{Kiran}~\snm{Lakshmipathaiah}}
\author[addressref=ursc]{\fnm{Ankur}~\snm{Kushwaha}\orcid{0000-0003-0034-1027}}
\author[addressref=ursc]{\fnm{M.~C.}~\snm{Ramadevi}}
\author[addressref=ursc]{\fnm{Smrati}~\snm{Verma}}

\address[id=ursc]{U.R. Rao Satellite Centre, Indian Space Research Organisation, Old Airport Road, Vimanapura PO Bengaluru, 560017, India}
\address[id=iitg]{Department of Physics, Indian Institute of Technology Guwahati, Guwahati, 781039, Assam, India}
\address[id=cessi]{Center of Excellence in Space Sciences India, Indian Institute of Science Education and Research Kolkata, Kolkata, 741246, India}

\runningauthor{Sarwade et al.}
\runningtitle{Iron Fluorescence in X-class Solar Flares}

\begin{abstract}
Iron fluorescence is produced by the irradiation of the solar photosphere by coronal X-rays during flares.
This study presents the first comprehensive analysis of iron K$\alpha$ fluorescence characteristics in 47 X-class flares observed during the inaugural year of the Solar Low Energy X-ray Spectrometer (SoLEXS) on board India's Aditya-L1 mission.
Leveraging the capability of modern silicon drift detectors (SDDs) for simultaneous broadband continuum and line measurements, the Fe K$\alpha$ flux and the exciting flux ($F_{>7.11 \text{ keV}}$) are quantified for each event, establishing a well-determined relationship between them across the sample.
The derived fluorescence efficiencies exhibit a center-to-limb dependence consistent with theoretical models, offering a potential diagnostic to probe coronal source heights and viewing geometries.
While statistical uncertainties currently limit the ability to track rapid height variations on short timescales, the mean fluorescence efficiency during the flare peak provides a potential constraint on the effective coronal source height. However, this derivation remains subject to a fundamental degeneracy, as the estimated source height cannot be uniquely determined without assuming a specific value for the photospheric iron abundance.
These findings demonstrate that SDDs, despite having lower spectral resolution than traditional crystal spectrometers, provide a new diagnostic for the solar iron fluorescence observations.
\end{abstract}


\end{frontmatter}

\section{Introduction}
The solar corona, characterized by plasma temperatures exceeding several million Kelvin, emits strongly across the soft X-ray and extreme ultraviolet spectra. This emission arises from optically thin plasma dominated by free-free (bremsstrahlung) continuum and line transitions in highly ionized elements \citep{benzFlareObservations2017,delzannaSolarUVXray2018}.
During solar flares, where a large amount of energy (up to \SI{1e32}{\erg}) is released over short timescales of a few minutes, magnetic reconfiguration drives plasma heating, which in turn produces intense soft X-ray emission \citep{Shibata2011}.
Soft X-ray spectra indicate that electron temperatures can reach \SI{\sim 30}{\mega \kelvin} and potentially up to \SI{\sim 50}{\mega \kelvin} in super-hot flare cores \citep{Caspi2014, Warmuth2016}, with emission measures up to \SI{1e50}{\centi \metre \tothe{-3}} during large flares \citep{Phillips2008}.

A fraction of the X-rays from these high-temperature coronal sources irradiates the cooler, denser photosphere below.
When incident X-ray photons have energies exceeding the K-shell binding energy of abundant metals like iron (\SI{7.11}{\keV}), they can induce inner-shell ionization.
The subsequent fluorescent emission occurs as outer-shell electrons fill these vacancies, producing characteristic fluorescent line radiation.
The most prominent photospheric fluorescence feature is the Fe K$\alpha$ doublet at \SI{6.4}{\keV} (\SI{1.94}{\angstrom}), resulting from 1s-2p transitions in neutral or mildly ionized iron.
This fluorescence mechanism was first proposed by \cite{Neupert1967} and later quantified through Monte Carlo simulations by \cite{Basko1978,Basko1979} and \cite{Bai1979}.

Early observational validation of photospheric fluorescence came from high-resolution crystal spectrometers.
The Bent Crystal Spectrometer on the Solar Maximum Mission (SMM) \citep{ActonSMMXRP1980} resolved the Fe K$\alpha1$ and K$\alpha2$ lines in \num{40} large flares \citep{Parmar1984}.
The Bragg Crystal Spectrometer on Yohkoh \citep{CulhaneYohkohBCS1991} detected the Fe~K$\beta$ line at \SI{7.06}{\keV} in \num{27} flares, further confirming analogous fluorescence processes \citep{PhillipsIronKBeta1994}.
While these crystal spectrometers achieved exceptional spectral resolution ($\Delta$E/E \num{\sim 1e-3} or $\Delta$E \SI{\sim 10}{\electronvolt}) \citep{Phillips2004b}, they operated within narrow energy passbands optimized for specific line diagnostics.

Consequently, these instruments could not simultaneously measure the broadband exciting flux above \SI{7.11}{\keV} alongside the fluorescence lines. This limitation prevented a direct determination of the fluorescence efficiency, making it difficult to constrain the excitation mechanism and fully validate theoretical predictions.

Over the past decade, observational priorities have shifted toward broadband (\SIrange{1}{20}{\keV}) spectroscopy using silicon drift detectors (SDDs) \citep{Niculae2006}.
Instruments such as Miniature X-ray Solar Spectrometer (MinXSS) \citep{MooreMINXSS2018}, Solar X-ray Monitor (XSM) \citep{Mithun2020}, Dual Aperture X-ray Solar Spectrometer (DAXSS) \citep{SchwabDAXSS2020}, etc. have demonstrated the advantages of SDDs, which offer continuous spectral coverage and superior count rate handling capabilities at a moderate resolution (\SI{\approx 170}{\electronvolt} at \SI{6}{\keV}).
This broadband capability enables simultaneous measurement of emission lines and continuum emissions, facilitating diagnostics of plasma properties such as temperature, emission measure (EM), and elemental abundances through thermal spectral modeling.
Even with SDDs' moderate resolution, spectral fitting residuals near \SI{6.4}{\keV} (Fe K$\alpha$) reveal systematic deviations (e.g., \citet[Figure 6]{mithunSoftXRaySpectral2022}, \citet[Figure 3]{Ng2025}), indicating possible unaccounted fluorescence contributions, which have been attributed to iron fluorescence in subsequent analyses \citep{MithunThesis}.

Solar Low Energy X-ray Spectrometer (SoLEXS) \citep{Sankarasubramanian2025} on board Aditya-L1, is one of the SDD-based soft X-ray spectrometers.
This study presents the first systematic analysis of iron fluorescence using SoLEXS, focusing on \num{47} X-class flares observed during its inaugural operational year in 2024. The analysis is organized as follows: Section~\ref{sect:fluo_mechanism} describes iron fluorescence mechanism; Section~\ref{sect:instrument} details SoLEXS design and in-flight performance; Section~\ref{sect:data_selection} outlines X-class flare selection criteria; Section~\ref{sect:data_analysis} presents the spectral fitting methodology for quantifying Fe K$\alpha$ line flux and exciting flux above \SI{7.11}{\keV}; Section~\ref{sect:fluo_characteristic} analyzes the temporal evolution of the Fe K$\alpha$ flux and the Fe K$\alpha$-to-exciting flux ratio as a function of heliocentric angle, comparing observations with theoretical predictions.

\section{The Iron Fluorescence Mechanism} \label{sect:fluo_mechanism}
\begin{figure}
    \centering
    \includegraphics[width=0.85\linewidth]{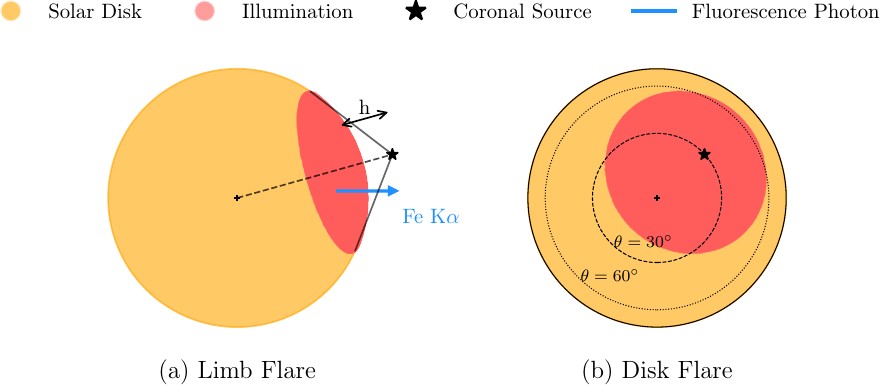}
\caption{Schematic illustrating the geometry of photospheric iron fluorescence excited by a coronal X-ray source.
(a) Limb flare configuration: The coronal source (black star), located at a height $h$ above the photosphere, illuminates a spherical cap on the solar surface (highlighted in red).
This illuminated region is the site of fluorescence production. A blue arrow indicates the escaping Fe K$\alpha$ fluorescent photons.
Tangent lines illustrate the horizon visible from the source, defining the maximum extent of the illuminated area capable of producing fluorescence.
(b) Disk flare configuration: The coronal source is positioned at a heliocentric angle of $\theta = $ \ang{30}. The corresponding illuminated photospheric area (red) is shown projected onto the solar disk.
The solid angle subtended by the photosphere at the source determines the fraction of X-rays intercepted and thus directly proportional to the total fluorescence efficiency.}

    \label{fig:fluorescence_schematic}
\end{figure}

X-ray sources in solar flares have been observed at coronal heights exceeding \SI{200}{\mega\metre} (\num{0.3} R$_\odot$) above the photosphere \citep{Kane1992, Krucker2008}.
While most of this radiation escapes into space, a fraction is directed downward, illuminating the cooler, denser layers of the solar atmosphere below (Figure~\ref{fig:fluorescence_schematic}). This interaction gives rise to a distinct set of spectral signatures, among which fluorescence lines of neutral or weakly ionized iron are present.

The iron fluorescence is produced when X-ray photons with energies above the K-shell binding energy of neutral iron (\SI{7.11}{\keV}) penetrate the chromosphere and are absorbed by photospheric iron atoms. Photoionization ejects a 1s electron, creating a vacancy in the innermost shell. As an electron from an outer shell, typically the L-shell (n=2), fills this vacancy, the atom relaxes and emits a characteristic Fe K$\alpha$ line at \SI{6.40}{\keV}. A less probable transition from the M-shell (n=3) produces the weaker Fe K$\beta$ line at approximately \SI{7.06}{\keV}.

A comprehensive quantitative framework for this X-ray fluorescence process was first developed by \cite{Bai1979} using Monte Carlo simulations. The model identifies key parameters that together determine the flux of fluorescent photons escaping the photosphere and observed at some distance. The Fe K$\alpha$ flux ($F_\mathrm{K\alpha}$) can be expressed as:

\begin{equation}
    F_{K\alpha}(\theta) = \Gamma(T,A_{Fe},h) \, f(\theta,h) \cdot \frac{F_{>7.11}}{1 + \alpha(T, h, \theta)}.
\end{equation}

\begin{figure}[htbp]
    \centering
    \centerline{
        \includegraphics[width=0.32\textwidth,clip=]{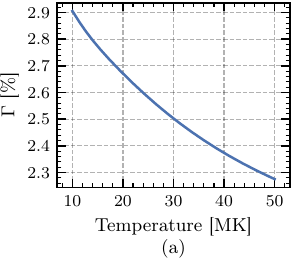}%
        \hspace*{0.005\textwidth}
        \includegraphics[width=0.32\textwidth,clip=]{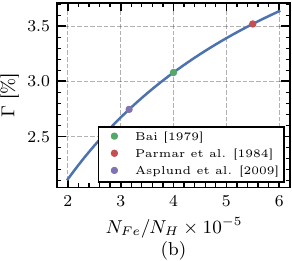}%
        \hspace*{0.005\textwidth}
        \includegraphics[width=0.32\textwidth,clip=]{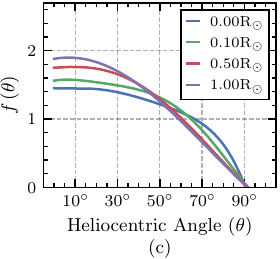}%
    }

    \caption{
        (a) The intrinsic fluorescence efficiency ($\Gamma$) as a function of coronal plasma temperature, showing that hotter flares are less efficient at producing fluorescence.
        (b) The dependence of the intrinsic fluorescence efficiency ($\Gamma$) on the photospheric iron abundance, which shows a slightly sub-linear scaling.
        (c) The geometric factor ($f(\theta,h)$) as a function of heliocentric angle ($\theta$) for several different coronal source heights (\textit{h}), based on the parameterization by \citet{Drake2008}.
    }
    \label{fig:fluo_depen}
\end{figure}

Here, $\Gamma(T,A_{Fe},h)$ represents the intrinsic fluorescence efficiency, defined as the fraction of incident exciting X-ray photons that successfully produce an escaping Fe K$\alpha$ photon.
The terms in this equation encapsulate the fundamental physical dependencies:
\begin{itemize}
    \item Exciting Photon Flux ($F_{>7.11}$): This term represents the total flux of coronal X-ray photons with energies above the \SI{7.11}{\keV} K-shell ionization threshold of neutral iron.
Although non-thermal emission may contribute to this band, the spectrum in the \SIrange{7}{12}{\keV} range for X-class flares is dominated by thermal emission.
At the \SI{7.11}{\keV} Fe K-edge energy, the typical height of absorption is deep in the solar atmosphere, approximately \SI{100}{\kilo\metre} above the $\tau_{5000}=1$ level of the photosphere \citep{Drake2008}.

    \item Coronal Source Temperature ($T$):
    The intrinsic fluorescence efficiency, $\Gamma$, decreases as the plasma temperature increases (Figure~\ref{fig:fluo_depen}a). This occurs because hotter plasmas produce a harder X-ray spectrum, and higher-energy photons are more likely to be Compton scattered rather than absorbed via photoionization, whose cross-section falls off rapidly with energy.

    \item Photospheric Iron Abundance ($A_{Fe}$):
    The intrinsic fluorescence efficiency, $\Gamma$, also scales with the number of iron atoms available for fluorescence, though the relationship is slightly sub-linear (Figure~\ref{fig:fluo_depen}b). The precise value for the photospheric abundance has been refined over time, with modern 3D models yielding values such as \SI{3.16e-5}{} \citep{Asplund2009}.

    \item Heliocentric Angle ($\theta$):
    This geometric factor, modeled by the function $f(\theta,h)$, determines the optical depth along the observer's line of sight. The observed flux is highest for disk-center flares ($\theta = $ \ang{0}) and is suppressed toward the limb ($\theta = $ \ang{90}) due to increased absorption.
    The angular dependence function, $f(\theta,h)$, has been conveniently fitted with fifth-order polynomials by \citet{Drake2008}, which are employed in this work (Figure~\ref{fig:fluo_depen}c).

\item Coronal Source Height ($h$):
The altitude of the X-ray source alters both the intrinsic efficiency $\Gamma$ and the angular distribution $f(\theta,h)$. The height dependence of $\Gamma$ is primarily caused by the variation in solid angle subtended by the photosphere at different flare heights. Thus, a flare which occurs at $h$=\SI{0}{} will produce about twice as much K$\alpha$ flux as one at $h$=\SI{0.1}{R_{\odot}}.
Furthermore, the decrease in $\Gamma$ is steeper than a strictly proportional relation to the solid angle.
As $h$ increases, the incoming X-ray photons strike the photosphere more vertically. Consequently, they penetrate deeper into the atmosphere on average, making it less likely for the resulting K$\alpha$ photons to escape the photosphere. However, because these deeply generated photons can successfully escape only if they are traveling nearly radially outward, the emergent emission becomes beamed
in the vertical direction. For flares near disk center ($\theta \approx 0^\circ$), this beaming is directed toward the observer, causing the normalized angular factor, $f(0,h)$, to actually increase with height. Conversely, for flares near the limb (large $\theta$), this vertical beaming directs photons away from the observer's line of sight, increasing the absorption and leading to a steeper decrease in the observed flux.
The relative height dependence of $\Gamma$, employed in this work, is also borrowed from \citet[Table 2]{Drake2008}.

    \item Albedo Correction ($\alpha$):
    This term accounts for the small fraction (\SI{\approx 4}{\percent}) of X-rays that are Compton back-scattered from the photosphere \citep{Bai1979}. This minor correction is neglected for this study.

\end{itemize}

    Although alternative excitation mechanisms, such as direct impact by non-thermal electrons spiralling down from the corona, have been suggested \citep{Acton1965, Phillips1973}, their contribution is physically constrained by the electron stopping depth. Standard thick-target models and radiative hydrodynamic simulations indicate that the bulk of non-thermal electrons (typically tens of keV) undergo Coulomb collisions and deposit their energy in the chromosphere \citep{Brown1971, Emslie1978, Allred2015}. Only in rare, intense events do high-energy electrons possess sufficient energy to penetrate through the chromosphere and reach the photospheric layers, a phenomenon typically associated with White Light Flares \citep{Hudson2006, MartnezOliveros2012, Kowalski2017}. Consequently, while collisional excitation may contribute during the impulsive phase, it is unlikely to be significant during the gradual thermal phase. This is supported by the time evolution of the Fe K$\alpha$ flux, which closely follows the thermal soft X-ray flux (E\SI{>7.11}{\keV}) rather than the impulsive hard X-ray bursts, as established by early crystal spectrometer observations \citep{Parmar1984, Zarro1992, PhillipsIronKBeta1994}. More recently, using modern SDD-based spectroscopy, \citet{MithunThesis} corroborated this by showing that the Fe K$\alpha$ line excess tracks the exciting thermal X-ray flux while showing no significant correlation with the non-thermal electron flux. Collectively, this evidence favors the X-ray photon fluorescent excitation model over electron collisional excitation mechanisms for the bulk of the thermal flare phase.

This study exploits the unique capabilities of SoLEXS to simultaneously measure both $F_{K\alpha}$ and $F_{>7.11}$, permitting the most direct application of this theoretical X-ray fluorescence framework.

\section{SoLEXS Instrument on Board Aditya-L1} \label{sect:instrument}

Aditya-L1 is India's first space-based solar mission developed by the Indian Space Research Organisation (ISRO). Launched on 2 September 2023, the spacecraft was inserted in a halo orbit around Lagrangian point 1 (L1) of the Sun-Earth system on 6 January 2024.
This strategic vantage point provides a continuous, uninterrupted view of the Sun, eliminating the eclipses and South Atlantic Anomaly transits that limit the effective observing time of low-Earth-orbit missions.

SoLEXS, on board Aditya-L1, is a Sun-as-a-star soft X-ray spectrometer designed for continuous monitoring of solar soft X-ray activity.
It operates across the \SIrange{2}{22}{\keV} energy range with a high count-rate capability (\SI{\sim 1e5}{\counts \per \second}) and an energy resolution of approximately \SI{170}{\electronvolt} (FWHM) at \SI{5.9}{\keV}.
SoLEXS combines the advantages of SDD technology with uninterrupted solar observations from L1, enabling unprecedented continuous monitoring of solar activity from A-class to X-class flares.

To accommodate this vast dynamic range of solar X-ray emission, SoLEXS utilizes two identical SDDs with different aperture areas.
SDD1, with an aperture area of \SI{7.106}{\milli \metre \squared}, is designed for weak flares (A-, B-, and C-class), while SDD2, with an aperture area of \SI{0.106}{\milli \metre \squared}, is used to observe intense C-, M-, and X-class flares without saturation.
The cylindrical apertures provide fields of view of \SI{\pm 1.8}{\degree} and \SI{\pm 1.3}{\degree} for SDD1 and SDD2, respectively.
Both detectors have an active area of \SI{30}{\milli \metre \squared}, a thickness of \SI{450}{\micro\metre}, and an \SI{8}{\micro\metre} DuraBeryllium Plus entrance window.

The SDDs are maintained at an operating temperature of approximately \SI{-45}{\celsius} by integrated two-stage Peltier coolers. Heat is efficiently extracted via copper thermal flanges connected to external aluminum radiators, a process that minimizes thermal noise and preserves the instrument's spectral resolution.
The on board Digital Pulse Processing (DPP) system converts analog SDD signals into energy spectra at a 1-second cadence.
This system uses an adaptive energy binning scheme to provide better channel width (\SI{47.75}{\electronvolt \per \channel}) below \SI{8}{\keV}, which contains the majority of solar coronal emission lines \citep{Phillips2004}, while grouping the subsequent channels into pairs, doubling the channel width (94.5 eV/channel) at higher energies. The DPP also incorporates pile-up rejection logic that enables high count-rate operation without spectral degradation.

SoLEXS commenced solar observations on 13 December 2023 and has been continuously monitoring solar soft X-rays with a nearly \SI{100}{\percent} duty cycle.
The continuous, high-cadence spectral data provide a rich dataset for systematic studies of flare energetics and coronal heating mechanisms. A detailed description of the instrument is available in \citet{Sankarasubramanian2025}. Ground and in-flight calibration details are provided in \cite{Sarwade2025}.

\subsection{On-Board Calibration Source and Background Spectrum}
SoLEXS is equipped with an internal $^{55}$Fe radioactive calibration source to ensure stable performance and accurate energy calibration throughout the mission.
The source undergoes electron capture decay to $^{55}$Mn, producing characteristic Mn K$\alpha$ (\SI{5.89}{\keV}) and K$\beta$ (\SI{6.49}{\keV}) X-ray lines. The \SI{6}{\micro\metre} titanium foil generates additional Ti fluorescence at \SI{4.51}{\keV} (K$\alpha$) and \SI{4.93}{\keV} (K$\beta$), providing multiple reference energies. Emitting at a very low count rate, this source provides continuous in-flight calibration without contaminating the solar spectrum.

\begin{figure}
    \centering
    \includegraphics{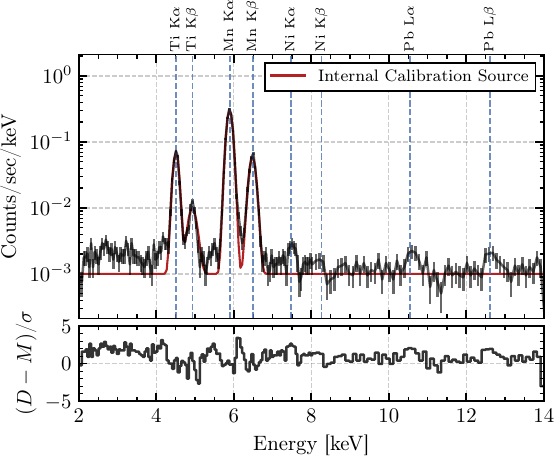}

    \caption{
        In-flight calibration spectrum and background characterization.
        The plot shows the spectrum recorded by SDD2 during the pre-commissioning phase with the aperture door closed, integrated over \SI{48}{hours}. The observed spectrum (black points) is dominated by characteristic lines from the on board $^{55}$Fe calibration source, including Ti K$\alpha$, Ti K$\beta$, Mn K$\alpha$, and Mn K$\beta$ (fitted model in solid red). Several weaker fluorescence lines from instrumental materials such as Ni and Pb are also identified.
        The lower panel shows the residuals after subtracting the best-fit model of the calibration and known instrumental lines. The absence of any significant residual features near \SI{6.40}{\keV} and \SI{7.06}{\keV} confirms that no detectable Fe K$\alpha$ or Fe K$\beta$ fluorescence is generated within the instrument itself.
    }
    \label{fig:bg_spec}
\end{figure}

The instrument's intrinsic background needs to be characterized to validate the solar origin of spectral features.
This was accomplished during the pre-commissioning phase by acquiring long-exposure spectra with the closed aperture door.
The resulting background spectrum reveals characteristic fluorescence lines originating from the instrument's structural components. Characteristic lines corresponding to nickel and lead
are observed in the background spectrum (Figure~\ref{fig:bg_spec}).

The background spectra confirm that no significant iron fluorescence originates from within the instrument itself.
The Mn K$\beta$ line (\SI{6.49}{\keV}) from the calibration source lies close in energy to the Fe K$\alpha$ line (\SI{6.40}{\keV}), but spectral fitting of the aperture closed condition spectrum shows no significant residuals when modeled with a Gaussian centered at Mn K$\beta$, indicating the lack of Fe K$\alpha$ contribution. Moreover, no detectable emission is observed at the Fe k$\beta$ line (\SI{7.06}{\keV}), which would arise along with Fe K$\alpha$ if fluorescence occurred in instrumental materials.
This finding is consistent with pre-launch ground calibration tests using a copper X-ray gun, where no Fe fluorescence lines were detected, even when the instrument aperture was open and illuminated by intense X-rays.
This combined absence indicates that iron fluorescence does not occur within the instrument itself, and any Fe K$\alpha$ and K$\beta$ features observed during solar observations are of solar origin.

\subsection{Spectral Resolution and Line Detection Capabilities}
SoLEXS SDD2 detector provides a spectral resolution of \SI{171.2 \pm 5.9}{\electronvolt} (FWHM) at \SI{5.9}{\keV}, with an energy binning of $47.52 \pm 0.33$ eV/channel below \SI{8}{\keV}.
While this moderate resolution does not resolve the individual, closely-spaced lines within the emission complexes of highly ionized elements (e.g., S, Ar, Ca, Fe, Ni \citep{Phillips2004}), it is sufficient to distinguish the major line features from the underlying thermal bremsstrahlung continuum.

The soft X-ray spectrum of flares contains two diagnostically important line groups. The first, often called the ``Fe line feature'', is a prominent complex at \SI{\approx 6.7}{\keV}. It is a blend of the Fe {\sc xxv} resonance line at \SI{6.7}{\keV} and numerous associated dielectronic satellite lines.
A secondary, weaker emission feature is also detectable near \SI{\approx 8}{\keV}, referred to as the ``Fe/Ni line feature'', which is a composite of iron transitions (Fe {\sc xxv} and Fe {\sc xxvi}) and lines from highly ionized nickel (Ni {\sc xxvii} and Ni {\sc xxviii}) \citep{Phillips2004}.

\begin{figure}[htbp]
    \centering
    \includegraphics[width=0.49\textwidth]{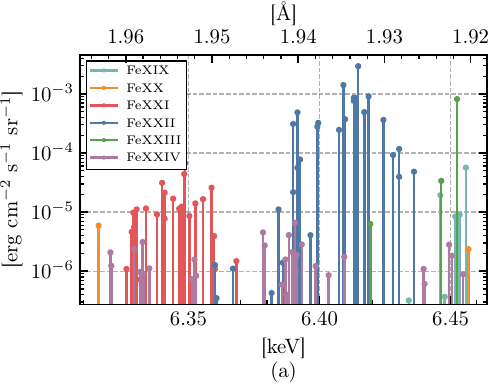}
    \includegraphics[width=0.49\textwidth,trim={-0.30in 0.in 0.0in 0.in},clip=]{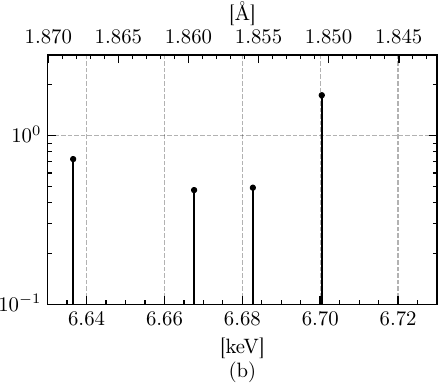}
    \caption{Simulated line intensities using the CHIANTI database for a plasma at \SI{20}{\mega\kelvin}.
    (a) The region around the neutral Fe K$\alpha$ energy (\SI{6.40}{\keV}), showing the weak contributions from lower ionization states (Fe {\sc xix}-Fe {\sc xxiv}).
    (b) The region around the main thermal Fe line complex (\SI{6.7}{\keV}), showing the dominant high-temperature lines (Fe {\sc xxv}). Note the difference in intensity scales, illustrating that the lines near \SI{6.4}{\keV} are negligible compared to the main complex at this temperature.}
    \label{fig:fe_line_simulations}
\end{figure}

The composition of the Fe line feature is highly dependent on the plasma temperature, which is a key factor in differentiating the Fe K$\alpha$ fluorescence at \SI{6.40}{\keV}. At the high temperatures characteristic of X-class flares (\SI{\approx 20}{\mega\kelvin} or higher), the feature's emission is overwhelmingly dominated by lines from the highest ionization stage, clustered between \SI{6.6}{\keV} and \SI{6.7}{\keV} (Figure~\ref{fig:fe_line_simulations}b). The satellite lines from lower ionization stages (Fe {\sc xix--xxiv}), which lie closer in energy to the fluorescence line, are significantly weaker at these temperatures \citep[Figure~1]{Phillips2004}.

To visualize potential contamination from lower-ionization species, the emission lines spectrum at \SI{20}{\mega\kelvin} is modeled using the CHIANTI atomic database \citep[Version 11]{chianti11}. As shown in Figure~\ref{fig:fe_line_simulations}a, a large number of faint satellite lines (primarily from Fe {\sc xix--xxiv}) are indeed present in the immediate vicinity of \SI{6.4}{\keV}. Although these lines fall within the instrumental FWHM of the neutral Fe K$\alpha$ feature, their intensity is orders of magnitude lower than the main thermal complex. Furthermore, the ionization fraction of these states is at least one order of magnitude lower than that of Fe {\sc xxv} at \SI{20}{\mega\kelvin} \citep{Gu2003}.
While these factors suggest that the satellite contribution is likely faint, a detailed quantitative assessment is presented in Section~\ref{low_temp_contri} to demonstrate that these lines are negligible compared to the observed fluorescence signal.

In contrast to these faint overlapping lines, the strong thermal Fe line complex at $\approx$ \SI{6.7}{\keV} is handled by the instrument's spectral resolution.
The energy difference of $\approx$ \SI{300}{\electronvolt} between the fluorescence line (\SI{6.4}{\keV}) and the thermal peak is nearly twice the instrumental FWHM of SoLEXS. This separation, combined with an energy binning that provides \SIrange{6}{7}{} spectral channels across this range, ensures that the fluorescent feature can be clearly distinguished from the thermal continuum and the hot Fe line complex during spectral fitting.

\section{Data Selection} \label{sect:data_selection}
The dataset for this study is compiled from observations made by SoLEXS during its first year of science operations in 2024. This period coincided with the exceptionally active maximum of Solar Cycle \num{25}. The year 2024 was one of the most productive periods for major solar flares, yielding a total of \num{54} X-class events. This provided an unprecedented opportunity to study iron fluorescence with high-cadence, broadband spectral data.

The selection is done by the following criteria:

\begin{itemize}
    \item Flare Class: Only X-class flares are chosen to ensure a high signal-to-noise ratio for the relatively weak Fe K$\alpha$ fluorescence line.
    \item Detector and Data Quality: Due to the high photon flux from these intense events, the SDD1 detector was frequently saturated \citep{Sarwade2025}. Therefore, all data used in this analysis are exclusively from the SDD2 detector, which is specifically designed with a smaller aperture to observe intense flares without saturation.
    \item Valid Time Intervals: SoLEXS Level-1 data are filtered using the corresponding Good Time Interval (GTI) files to ensure all instrument health parameters were nominal and the Sun was correctly positioned within the instrument's field of view.
\end{itemize}

This process yielded a final sample of \num{47} X-class flares (Table~\ref{tab:flare_list}). The dataset was significantly enhanced by periods of intense solar activity in May and October 2024. May 2024 was the most active month of Solar Cycle \num{25}.
In particular, the super-active region NOAA 13664 produced \num{12} X-class flares \citep{Li2024}, covering a broad range of heliocentric angles from $\theta \approx$ \ang{25} to the limb at \ang{90}. Similarly, NOAA 13663 contributed \num{5} X-class flares ranging from $\theta \approx$ \ang{25} to \ang{60}.
October 2024 featured several major events as well, including the X9.0 flare on 3 October, the strongest of the cycle to date.

Flare locations are obtained from the Solar Region Summary (SRS) data product \footnote{\url{https://www.swpc.noaa.gov/products/solar-region-summary}} to calculate the heliocentric angle ($\theta$) for each event.
The resulting sample provides a comprehensive distribution across the solar disk: approximately \SI{28}{\percent} of the flares (\num{13} events) are located near the disk center ($\theta \le$ \ang{30}), \SI{32}{\percent} (\num{15} events) are in the \ang{30} $< \theta <$ \ang{60} range, and \SI{40}{\percent} (\num{19} events) are near the limb (\ang{60} $\le \theta \le$ \ang{90}). This broad coverage, also illustrated in Figure~\ref{fig:solexs_fluo_all}, is essential for investigating the center-to-limb dependence of the fluorescence efficiency.

\begin{figure}[htbp]
    \centering

    \centerline{
        \includegraphics[width=0.45\textwidth,trim={0.15in 0.2in 0.15in 0.2in},clip=]{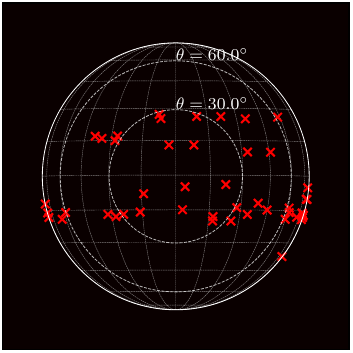}%
        \hspace*{0.03\textwidth}
        \includegraphics[width=0.49\textwidth,clip=]{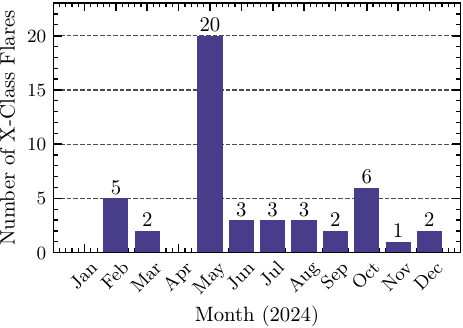}%
    }

    \vspace{-0.41\textwidth}
    \centerline{\large \bf
        \hspace{0.01 \textwidth} \color{white}{(a)}
        \hspace{0.5\textwidth} \color{black}{(b)} %
       \hfill}

    \vspace{0.38\textwidth} %

    \caption{(a) Spatial distribution of the X-class solar flares selected for this study. The heliographic latitude and longitude of each of the \num{47} flares in our sample are shown as red data points. The solid circle represents the solar limb ($\theta = \SI{90}{\degree}$), with the inner dashed circles representing contours corresponding to heliocentric angles of \SI{30}{\degree} and \SI{60}{\degree}. The plot demonstrates the comprehensive coverage of the dataset, with events distributed across a wide range of heliocentric angles.
    (b) Monthly distribution of X-class solar flares observed by SoLEXS in \num{2024}. The bar chart displays the number of X-class flares recorded each month, corresponding to the \num{47} events selected for this study.}
    \label{fig:combined_flare_overview}
\end{figure}

\section{Data Analysis} \label{sect:data_analysis}

\subsection{Spectral Fitting Methodology}
The solar spectrum as measured by spectrometers such as SoLEXS does not directly represent the incident X-ray spectrum due to instrumental response that modifies the detected photon distribution.
The observed spectrum is modified by the detector's Ancillary Response Function (ARF) and Spectral Redistribution Function (SRF). The ARF quantifies the instrument's effective area, accounting for the geometric aperture size, the energy-dependent transmission of the DuraBeryllium Plus entrance window, and the quantum efficiency of the silicon detector. The SRF describes the probability distribution of recorded channels for a monoenergetic input photon (spectral redistribution). This function incorporates the primary Gaussian peak broadened by electronic and Fano noise, and the silicon escape peak \citep{Sarwade2025}.
Mathematically, the observed spectrum $C(I)$ is convolution of the incident photon spectrum $f(E)$, modified by the ARF $A(E)$, with SRF $R(I,E)$,

\begin{equation}
C(I)=\int f(E) A(E) R(I, E) d E,
\end{equation}

where $I$ is the instrument pulse height analysis (PHA) channel, and $E$ is the incident photon energy. The channel $I$ is related to the detected energy via a linear relationship ($E$ = Gain $\times$ $I$ + Offset) determined from ground and on board calibration \citep{Sarwade2025}.
This calibration allows the observed count spectrum $C(I)$ to be mapped to the energy space representation $C(E)$ used in subsequent calculations (Equation~\ref{eq:F_711}).

To determine $f(E)$ given observed spectrum $C(I)$, response $R(I, E)$, and effective area $A(E)$ requires inverting this equation. The inversion is not possible in general and, as such, tends to be non-unique and unstable to small changes in $C(I)$ \citep{xspec_guide}.
The alternative is using the forward fitting technique. Here, the model spectrum, $f(E)$, is described in terms of a few parameters.

The model spectrum is then convolved with the instrumental response and fitted with the observed spectrum to minimize the $\chi^2$ statistics. The resulting parameters are considered ``best-fit'' parameters, which best represent the source spectrum.
The Sherpa fitting application \citep{sherpa_416,sherpa_paper} is used for forward fitting the model spectrally to the observed spectrum to get the best-fit parameters.

To characterize the temporal evolution of the flare properties, the spectral fitting methodology is applied to the full duration of each event. The spectral data are integrated into \SI{10}{\second} time bins, a cadence chosen to ensure sufficient signal-to-noise ratio while preserving temporal resolution. This time-resolved analysis yields a sequence of best-fit parameters for each flare, the analysis of which is presented in Section~\ref{temporal_evolution}.

\subsection{Single-Temperature Isothermal Model} \label{sect:isothermal_model}

Solar flare spectra in the soft X-ray range are commonly modeled under the assumption of isothermal plasma emission, parameterized by temperature ($T$), emission measure ($EM$), and elemental abundances \citep{Culhane1969}.
While the emission measure is operationally approximated in many fitting codes as proportional to $n_e^2 V$ (assuming a uniform electron density $n_e$ over the emitting volume $V$, which is a sufficient assumption for spatially unresolved Sun-as-a-star measurements as used in this work), the rigorous definition involves the integral of the electron and emitting ion densities, $EM = \int n_e n_i dV$. The ion density $n_i$ is related to $n_e$ by the abundance of the element
and the temperature-dependent ion fraction derived from ionization equilibrium calculations.

In this work, spectral modeling is performed using the \texttt{f\_vth} model available in the \texttt{sunkit-spex}\footnote{\url{https://github.com/sunpy/sunkit-spex}} Python package. This model computes the total continuum component, encompassing thermal bremsstrahlung (free-free), free-bound radiative recombination, and two photon emission and derives line intensities using the CHIANTI atomic database \citep[Version 11]{chianti11}.
The pre-computed CHIANTI v11 databases utilized by the model in this analysis are publicly available\footnote{\url{https://github.com/ianan/fvth_stuff}}.
The \texttt{f\_vth} model, defined in the package, was modified locally to allow elemental abundances to vary as free parameters during fitting.
This adapted model is equivalent to \texttt{f\_vth\_abun} model \footnote{\url{https://hesperia.gsfc.nasa.gov/ssw/packages/xray/idl/f_vth_abun.pro}} in the Object Spectral Executive (OSPEX) package \citep{OSPEX_software}.

All spectral fits are performed over the energy range of \SIrange{2.2}{12}{\keV}.
The lower bound avoids known uncertainties in the instrument response below \SI{2.2}{\keV}, while the upper bound is chosen based on where the signal-to-noise ratio becomes low.
The plasma temperature, emission measure, and the abundances of Fe, Ca, Ar, and S are treated as free parameters.
A systematic error of \SI{4}{\percent} is added to the data to account for calibration uncertainties in the ARF and SRF.
This value is determined empirically and is consistent with the radiometric uncertainty of $\approx$ \SIrange{4}{5}{\percent} derived from cross-calibration with the Chandrayaan-2/XSM instrument \citep{Sarwade2025}.
Sensitivity testing confirms that variations in the derived physical parameters resulting from changes in the systematic error are well within the statistical uncertainties.

\begin{figure}[htbp]
    \centering

    \centerline{\hspace*{0.015\textwidth} %
        \includegraphics[width=0.46\textwidth,trim={0.25in 0.3in 0.24in 0.3in},clip=]{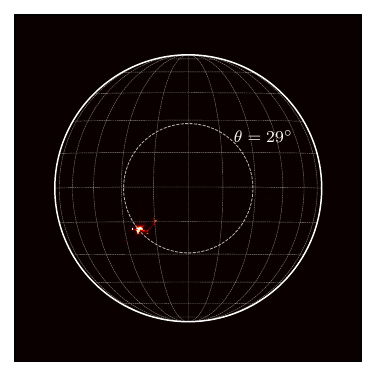}%
        \hspace*{0.03\textwidth}%
        \includegraphics[width=0.49\textwidth,clip=]{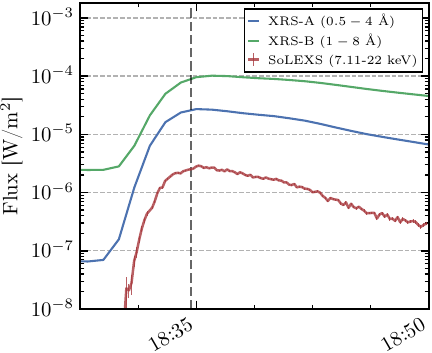}%
    }

    \vspace{-0.395\textwidth} %
    \centerline{\large \bf %
        \hspace{0.015 \textwidth} \color{white}{(a)} %
        \hspace{0.52\textwidth} \color{black}{(b)} %
       \hfill}

    \vspace{0.36\textwidth} %

    \centerline{
        \includegraphics[width=1.0\textwidth,clip=]{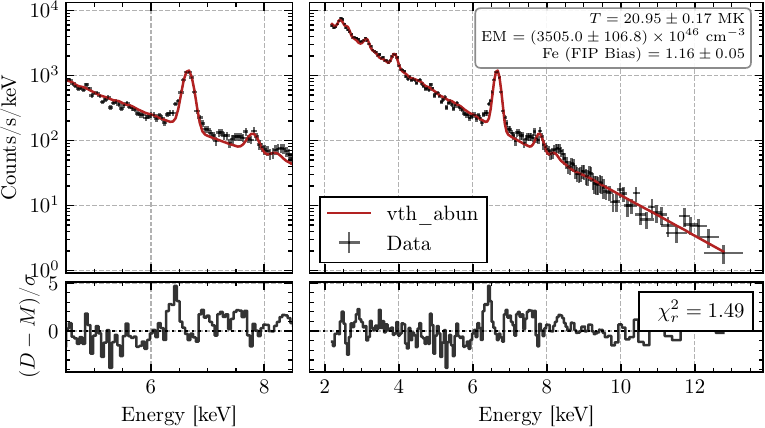}%
    }
    \vspace{-0.53\textwidth} %
    \centerline{\large \bf
    \hspace{.1 \textwidth} \color{black}{(c)} %
       \hfill}
    \vspace{0.49\textwidth} %

    \caption{
        (a) An AIA \SI{94}{\angstrom} difference image showing the flare's location at S17E24 (heliocentric angle $\theta \approx \SI{29}{\degree}$).
        (b) The corresponding GOES light curve for the X1.0-class flare (XRS-A and XRS-B) plotted alongside the SoLEXS \SIrange{7.11}{22}{\keV} flux (red).
        (c) The SoLEXS spectrum, integrated for \SI{10}{\second} starting at the flare's peak (SOL-2024-06-01T18:34:40, indicated by the dashed line in panel b). The right side of panel (c) shows the fit over the full energy range annotated with the best-fit temperature, emission measure, and Fe abundance, while the left side provides a zoomed-in view. A clear residual is observed near \SI{6.4}{\keV}, consistent with the presence of photospheric fluorescence for a near-disk-center event.
    }
    \label{fig:disk_flare_analysis_summary}
\end{figure}

\begin{figure}[htbp]
    \centering

    \centerline{\hspace*{0.015\textwidth} %
        \includegraphics[width=0.46\textwidth,trim={0.25in 0.3in 0.24in 0.3in},clip=]{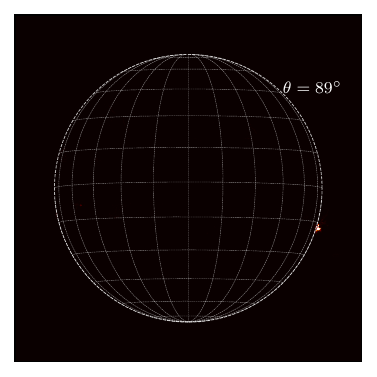}%
        \hspace*{0.03\textwidth}%
        \includegraphics[width=0.49\textwidth,clip=]{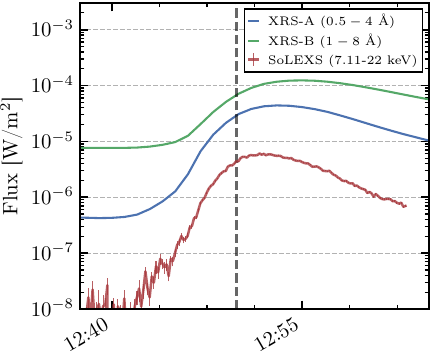}%
    }

    \vspace{-0.395\textwidth} %
    \centerline{\large \bf %
        \hspace{0.015 \textwidth} \color{white}{(a)} %
        \hspace{0.52\textwidth} \color{black}{(b)} %
       \hfill}

    \vspace{0.36\textwidth} %

    \centerline{
        \includegraphics[width=1.0\textwidth,clip=]{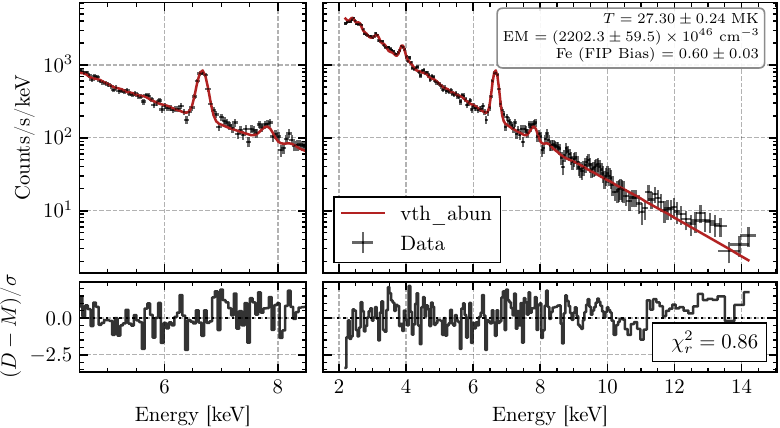}%
    }
    \vspace{-0.53\textwidth} %
    \centerline{\large \bf
    \hspace{.1 \textwidth} \color{black}{(c)} %
       \hfill}
    \vspace{0.49\textwidth} %

    \caption{
        (a) An AIA \SI{94}{\angstrom} difference image showing the flare's location at S17W89 (heliocentric angle $\theta \approx \SI{89}{\degree}$).
        (b) The corresponding GOES light curve for the X1.2-class flare (XRS-A and XRS-B) plotted alongside the SoLEXS \SIrange{7.11}{22}{\keV} flux (red).
        (c) The SoLEXS spectrum, integrated for \SI{10}{\second} starting at the flare's peak (SOL-2024-05-14T12:49:50, indicated by the dashed line in panel b). The right side of panel (c) shows the fit over the full energy range annotated with the best-fit temperature, emission measure, and Fe abundance, while the left side provides a zoomed-in view. In contrast to the near-disk-center event, no significant residual is observed near \SI{6.4}{\keV}, which is consistent with the suppression of photospheric fluorescence at the solar limb.
    }
    \label{fig:limb_flare_analysis_summary}
\end{figure}

Figures \ref{fig:disk_flare_analysis_summary}c and \ref{fig:limb_flare_analysis_summary}c present the results of this fitting procedure for two representative X-class flares.
Both flares are of similar magnitude of X1.0 and X1.2 (Figures~\ref{fig:disk_flare_analysis_summary}b~and~\ref{fig:limb_flare_analysis_summary}b) and exhibit a similar peak fitted temperature of approximately \num{22} and \SI{27}{\mega \kelvin}. The primary difference between them is their heliocentric angle. For the near disk-center flare ($\theta =$ \ang{29}, Figure~\ref{fig:disk_flare_analysis_summary}a), subtracting the best-fit isothermal model reveals residuals around \SI{6.4}{\keV} and \SI{7.06}{\keV}.
In contrast, for the limb flare ($\theta =$ \ang{89}, Figure~\ref{fig:limb_flare_analysis_summary}a), no such residuals are present. Since the flare class and plasma temperature are comparable, this pronounced difference in the spectral residuals can be attributed to their different viewing geometries.

Similar residuals at \SI{6.4}{\keV} have been reported in studies using more complex, multithermal models, which further bolsters the case for a secondary emission process separate from the hot coronal plasma.
For instance, \cite{mithunSoftXRaySpectral2022} found a persistent residual at \SI{6.4}{\keV} in a C8.3-class flare near disk-center at $\theta =$ \ang{18} even when using a multithermal differential emission measure (DEM) model. In fact, the residual became more prominent as more temperature components were added to the fit.
This excess was subsequently analyzed in detail by \citet{MithunThesis} using combined XSM and Spectrometer/Telescope for Imaging X-rays \citep[STIX]{Krucker2020_2} spectra, where it was modeled as Fe K$\alpha$ fluorescence. The measured line excess was found to closely track the exciting X-ray flux above the Fe K-edge, while showing no significant correlation with the non-thermal electron flux, consistent with a thermal fluorescence origin.
Similarly, in \cite{Ng2025}, a feature near \SI{6.4}{\keV} in an X2.5-class flare observed at $\theta =$ \ang{82} that is not accounted for by a two-temperature model can be seen.
Given the large heliocentric angle, this excess is unusually high for a standard X-ray fluorescence scenario.
While a detailed analysis of this rare phenomenon is reserved for a future study, the persistence of these residuals irrespective of thermal model complexity (as tested in Section~\ref{low_temp_contri}) strongly suggests they are not artefacts of an incomplete thermal description, particularly during the gradual phase.

The most plausible explanation for these features is photospheric X-ray fluorescence. The stark contrast between our disk and limb flare examples, and the consistent appearance of these residuals in other studies, aligns with the geometric dependence predicted by the fluorescence model of \cite{Bai1979}. The absence of these features in limb events suggests their origin in the photosphere, ruling out instrumental effects as a contributing mechanism.

\subsection{Low-Temperature Contribution}
\label{low_temp_contri}
A potential contaminating factor in identifying the \SI{6.4}{\kilo\electronvolt} fluorescence feature is the presence of satellite lines from lower ionization states of iron (Fe {\sc xix--xxiv}, Figure~\ref{fig:fe_line_simulations}a).
While the contribution of these lines at the fitted flare temperature is accounted for in the isothermal model described in Section~\ref{sect:isothermal_model}, additional flux could theoretically arise from a cooler plasma component (\SIrange{6}{10}{\mega\kelvin}).
Given that the spectral fitting range begins at \SI{2.2}{\kilo\electronvolt}, the broadband spectrum is dominated by the hot ($> \SI{15}{\mega\kelvin}$) flare component, potentially blinding the instrument to the continuum emission from cooler plasma \citep{Winebarger2012}.
Therefore, unmodeled emission lines from this ``hidden'' cool plasma, originating from the pre-flare background or a secondary cooler flaring component, could potentially contaminate the \SI{6.4}{\kilo\electronvolt} region.

To quantify the line contribution around \SI{6.4}{\kilo\electronvolt} due to this ``blind'' cooler plasma, a sensitivity analysis is performed using the CHIANTI atomic database (Version 11). The emission line spectrum is simulated for isothermal plasmas at various temperatures,
deliberately adopting higher values of emission measure and abundances to establish an upper limit of the potential contamination.
First, the EM value is fixed at \SI{1e51}{\per\cubic\centi\metre}. This value is chosen as a conservative upper limit to account for the high EM ratios (typically factors of \SIrange{3}{10}{}) often observed between cool and hot plasma components in multithermal flares \citep{Caspi2010, Woods2023}. It represents approximately \num{3} times the peak EM of the largest flare in our sample (X9.0 class, EM \SI{3.6e50}{\per\cubic\centi\metre}, Table~\ref{tab:flare_list}) and over \num{30} times the EM of the representative X1 flare (Figure~\ref{fig:disk_flare_analysis_summary}).
Second, the abundances are fixed to coronal abundance values.
While flaring plasma often exhibits abundances closer to photospheric values due to the injection of unfractionated material via chromospheric evaporation \citep{Warren2014,Doschek2018,mithunSoftXRaySpectral2022}, the background or pre-flare coronal plasma is subject to the First Ionization Potential (FIP) effect \citep{Laming2021}. The FIP effect typically enhances the abundances of low-FIP elements, such as iron, by a factor of \num{3} to \num{4} relative to the photosphere, yielding a maximum possible line intensity.

\begin{figure}[!h]
    \centering

    \includegraphics{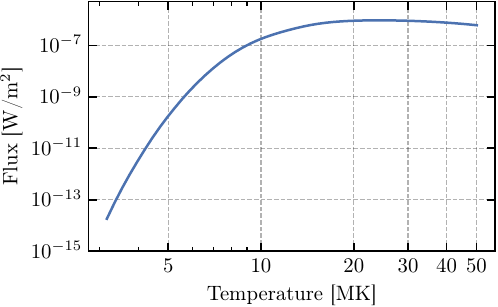}
    \caption{
        Simulated line flux in the SoLEXS spectral window centered at the Fe K$\alpha$ fluorescence energy (\SI{6.40 \pm 0.085}{\kilo\electronvolt}) as a function of plasma temperature. The simulation assumes an isothermal plasma with coronal abundances and a fixed emission measure of \SI{1e51}{\per\cubic\centi\metre}.
        }
    \label{fig:emission_line_flux_around_Fe_ka}
\end{figure}

The total line flux falling within the SoLEXS spectral resolution window centered at the fluorescence energy (\SI{6.40 \pm 0.085}{\kilo\electronvolt}) is then calculated as a function of temperature (Figure~\ref{fig:emission_line_flux_around_Fe_ka}).
The analysis reveals a strong dependence on temperature. At \SI{20}{\mega\kelvin} (typical for the hot flaring component), the simulated thermal line flux in this energy bin is $\sim \SI{1e-6}{\watt\per\square\metre}$.
However, an emission measure of \SI{1e51}{\per\cubic\centi\metre} at this temperature is physically extreme, exceeding the peak EM of even the largest reported historical flares (e.g., the X17 flare of SOL-2003-10-28 had an EM of \SI{1.3e50}{\per\cubic\centi\metre}; \citealt{Warmuth2016}).
Conversely, at cooler temperatures where such a high EM might be plausible (e.g., \SI{7}{\mega \kelvin}), the emissivity drops precipitously, and the line flux falls to $\sim \SI{1e-8}{\watt\per\square\metre}$.
This difference of two orders of magnitude demonstrates that even under upper-limit assumptions, the cool satellite lines are extremely faint in the \SI{6.4}{\kilo\electronvolt} band.

Furthermore, to empirically verify this theoretical limit of the impact of cooler plasma (whether pre-flare background or a secondary flaring component), a two-temperature component analysis is performed for the representative flare shown in Figure~\ref{fig:disk_flare_analysis_summary}. A pre-flare spectrum integrated over a \SI{3}{\minute} interval (18:26:00-18:29:00) is modeled with a single isothermal component (abundances fixed to coronal values), yielding a background temperature of \SI{6.52 \pm 0.86}{\mega\kelvin}, emission measure of \SI{1.77(0.09)e48}{\per \centi \metre \cubed}.
This background emission measure is three orders of magnitude lower than the \SI{1e51}{\per\cubic\centi\metre} value used in the flux simulations illustrated in Figure~\ref{fig:emission_line_flux_around_Fe_ka}, suggesting that the potential contamination from typical non-flaring background levels is minimal.

To test this, the peak flare spectrum (Figure~\ref{fig:disk_flare_analysis_summary}c) is fitted using a two-temperature model, where one component is fixed to these pre-flare background parameters.
The resulting best-fit values for the primary flare temperature and the residuals around \SI{6.4}{\keV} are consistent with those obtained from the single-temperature spectral modeling.
Even when the background emission measure is artificially increased by an order of magnitude, the fit parameters for the flaring component showed only marginal changes, and the residuals remained unaffected.
When the background emission measure is artificially fixed to \SI{1e50}{\per\cubic\centi\metre} (an order of magnitude lower than the simulation limit in Figure~\ref{fig:emission_line_flux_around_Fe_ka}), the spectral fit failed to converge.
Even at this reduced EM, the model overestimates the continuum flux in the \SIrange{2.2}{3}{\keV} range and drives the hot component temperature to \SI{\approx 28}{\mega\kelvin} (compared to the best fit isothermal temperature of \SI{20.95}{\mega\kelvin}), resulting in an overestimate at high energies (\SI{>9}{\keV}). Crucially, a residual at \SI{6.4}{\keV} persists.
This implies that any cool plasma component of such magnitude, whether background or flaring, is physically inconsistent with the observed spectrum, as it would no longer be obscured by the hot component.
This conclusion is independently corroborated by \citet{MithunThesis}, who demonstrated that even when explicitly employing a two-temperature model on XSM spectra extending down to \SI{1.3}{\keV}, where cooler plasma is better constrained, a residual at \SI{6.4}{\keV} persisted.
Consequently, cool-component modeling is deemed unnecessary for the full dataset, and the single-isothermal model is employed.

\subsection{Modeling the Fluorescence Contribution} \label{sect:composite_model}
The persistent residuals identified at \SI{6.4}{\keV} and \SI{7.06}{\keV} in on-disk flares necessitated an augmentation of the single-temperature thermal model. To account for the photospheric fluorescence, the model is expanded to include two empirical Gaussian profiles representing the Fe K$\alpha$ and K$\beta$ lines.

The parameters for these Gaussian components are constrained based on known atomic physics and instrumental characteristics.
The Fe K$\alpha$ component is centered at \SI{6.40}{\keV} with its width (FWHM) allowed to vary as a free parameter within the range of \SI{10}{\electronvolt} to \SI{300}{\electronvolt}.
Since the instrumental resolution (FWHM \SI{\approx 170}{\electronvolt}) is applied via convolution with the SRF, the Gaussian width parameter represents the pre-convolution profile.
This range is chosen to allow for intrinsic narrow lines (effectively delta functions) as well as potential instrumental line broadening effects that can occur at high count rates during the X-class flares.
While the Fe K$\beta$ component is centered at \SI{7.06}{\keV} with its width (FWHM) fixed at \SI{350}{\electronvolt}.
This FWHM was determined empirically to optimize fit convergence, as the weak K$\beta$ line prevents stable constraining of its width as a free parameter.
The flux ratio of the K$\beta$ to K$\alpha$ components is fixed at \num{0.134}, a value determined from the theoretical branching ratio and fluorescence yields for neutral iron \citep{BAMBYNEK1972}.

In addition to the Fe fluorescence, inspection of the residuals in high-flux spectra reveals systematic excesses near \SI{7.5}{\kilo\electronvolt} and \SI{8.3}{\kilo\electronvolt}. These features correspond to the characteristic Ni K$\alpha$ (\SI{7.48}{\kilo\electronvolt}) and Ni K$\beta$ (\SI{8.27}{\kilo\electronvolt}) energies. While these residuals are not consistently observed in all events, appearing more prominently in disk flare (Figure~\ref{fig:disk_flare_analysis_summary}c) compared to limb flare (Figure~\ref{fig:limb_flare_analysis_summary}c). Instrumental Ni fluorescence is a known background component, as identified in the pre-commissioning door-closed spectrum (Figure~\ref{fig:bg_spec}). To account for this signature, which may be excited by the intense solar X-ray flux illuminating the instrument structure, the spectral model is augmented with two additional Gaussian components fixed at the Ni fluorescence energies with widths of \SI{350}{\electronvolt}. Including these components effectively eliminates the residuals in the \SIrange{7}{8}{\kilo\electronvolt} range without affecting the parameters of the primary Fe K$\alpha$ feature.

\begin{figure}
    \centering

    \includegraphics[width=1.0\textwidth,clip=]{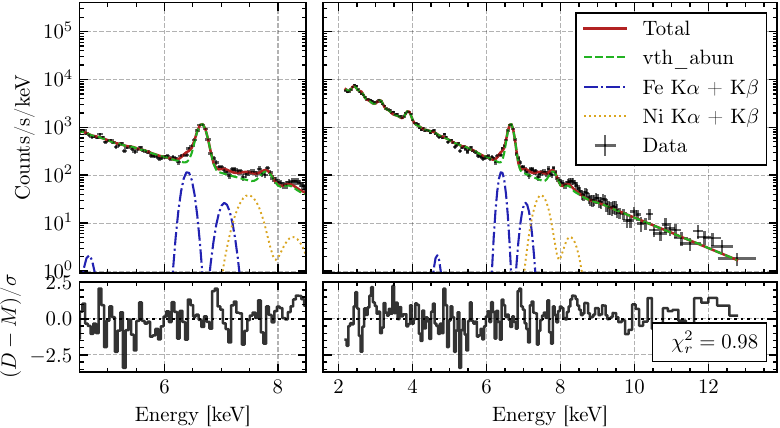}
    \caption{
        Composite model fit to the SoLEXS spectrum for a near-disk-center X1.0-class flare (location: S17E24, heliocentric angle: $\theta =$ \ang{29}).
        This spectrum is same as the one shown in Figure~\ref{fig:disk_flare_analysis_summary}c, integrated for \SI{10}{\second} starting at SOL-2024-06-01T18:34:40.
        The observed spectrum (black points) is shown with the total model fit (solid red line), which includes the thermal plasma component (vth\_abun, dashed green line) and the Fe K$\alpha$ + K$\beta$ fluorescence lines (dash-dot blue line) and the Ni K$\alpha$ + K$\beta$ instrumental fluorescence lines (dotted orange line).
        The best-fit thermal parameters are $T = \SI{20.44 \pm 0.18}{\mega \kelvin}$, $\text{EM} = \SI{3634.9 \pm 65.2 e46}{\per\cubic\centi\meter}$, and iron abundance (FIP bias) $= \num{1.19 \pm 0.05}$.
        The left panel provides a zoomed-in view.
        The inclusion of the fluorescence components improves the fit, accounting for the residuals.
        The weak Gaussian feature at \SI{\approx 4.66}{\keV} (dash-dot blue line) represents the instrumental silicon escape peak from the Fe K$\alpha$ line.
        }    \label{fig:disk_flare_isothermal_gaussian_fit}
\end{figure}

Combining the thermal component with the fluorescence Gaussians, this composite model allows for the simultaneous fitting of both the coronal plasma parameters (temperature, EM, and abundances) and the line-of-sight strength of the photospheric Fe fluorescence.
Figure~\ref{fig:disk_flare_isothermal_gaussian_fit} illustrates the application of this new model to the same near disk-center X1.0-class flare (Figure~\ref{fig:disk_flare_analysis_summary}). The inclusion of the Gaussian components provides a better fit to the data, accounting for the previous residuals. This is quantitatively confirmed by the improvement in the fit statistic, with the reduced $\chi^2$ value decreasing from \num{1.49} to \num{0.98}, indicating a much more accurate representation of the underlying physical processes.

\subsection{Quantifying the Fluorescence Flux}
With the composite spectral model successfully separating the thermal and fluorescence components, the flux of the Fe K$\alpha$ line is quantified by calculating the integrated flux under the best-fit Gaussian profile that represents the fluorescence feature.
The total photon flux, $F_{K\alpha}$ (in \SI{}{photons \per \centi \metre \squared \per \second}), is determined by the area of the Gaussian function:

\begin{equation}
F_{K\alpha} = \sqrt{2\pi} \cdot A_{K\alpha} \cdot \frac{1}{2\sqrt{2\ln 2}} \cdot \text{FWHM}_{K\alpha},
\end{equation}

where $A_{K\alpha}$ is the amplitude (in \SI{}{photons \per \centi \metre \squared \per \second \per \keV}) and $\text{FWHM}_{K\alpha}$ is the full width half maximum (in \SI{}{\keV}) of the fitted Gaussian. As described in the previous section, the centroid of the line is fixed at \SI{6.40}{\keV} while the width, $\text{FWHM}_{K\alpha}$, is allowed to vary as a free parameter in the fit, constrained to the range of \SI{10}{\electronvolt} to \SI{300}{\electronvolt}. Therefore, both the best-fit amplitude and width determine the line flux.
The statistical uncertainty in the flux is directly propagated from the \num{1}-$\sigma$ errors on both the fitted parameters, which are provided by the Sherpa fitting routine.

The K$\beta$ line is intrinsically weaker, with a flux of only \SI{\approx 13}{\percent} of the K$\alpha$ line, leading to larger relative uncertainties. Thus, only the flux of the stronger and more statistically robust Fe K$\alpha$ line is reported.

The photon flux is converted to an energy flux (in \SI{}{\watt \per \metre \squared}) by multiplying by the energy of each photon. For the Fe K$\alpha$ line, this conversion is:

\begin{equation}
F_{K\alpha} [\text{W m}^{-2}] = F_{K\alpha} [\text{photons }\text{cm}^{-2}\text{s}^{-1}]\times E_{K\alpha}\times 1.602\times 10^{-13},
\end{equation}

where $E_{K\alpha}$ is the line energy (\SI{6.4}{\keV}) and the constant converts the energy from \SI{}{\keV} to \SI{}{\joule} and the area from \SI{}{\centi \metre \squared} to \SI{}{\metre \squared}.

\subsection{Quantifying the Exciting Flux}
Having determined the flux of the fluorescent Fe K$\alpha$ line, the flux of the coronal X-rays responsible for fluorescence is computed next. According to the fluorescence model, this corresponds to the total photon flux (comprising both the thermal continuum and high-energy emission lines, such as the Fe/Ni complex) above the Fe K-edge energy of \SI{7.11}{\keV}. The ratio of the fluorescent line flux to this exciting flux provides a direct measure of the fluorescence efficiency.

The exciting flux, $F_{>7.11}$, is derived by directly integrating the observed counts in the spectrum, thereby avoiding assumptions about the thermal or non-thermal nature of the flare X-ray emission. The flux is calculated by summing the energy-weighted counts above the K-edge, adjusted for the instrument's effective area and exposure time:

\begin{equation}
\label{eq:F_711}
F_{>7.11} = \sum_{E=7.11}^{22} \frac{C(E) \cdot E}{A(E) \cdot t_{\text{exp}}} \quad [\text{W m}^{-2}].
\end{equation}

Here, $C(E)$ is the observed counts at energy $E$, $A(E)$ is the instrumental effective area, and $t_{\text{exp}}$ is the integration time. The summation extends to \SI{22}{\keV}, the upper energy limit of SoLEXS.
The statistical uncertainty on the integrated flux is calculated by propagating the Poisson and systematic errors from each energy channel included in the sum.
For this broadband flux integration, a diagonal response matrix is assumed. This is a reasonable simplification, as the energy redistribution effects described by the SRF have a minimal impact when summing over a wide energy range and are negligible compared to other sources of uncertainty.

\section{Fluorescence Characteristics} \label{sect:fluo_characteristic}
\subsection{Temporal Evolution and Model Comparison}
\label{temporal_evolution}
\begin{figure}
    \centering
    \includegraphics{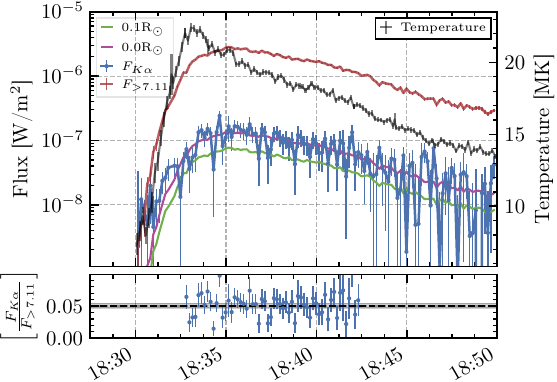}

\caption{Temporal evolution of observed and simulated fluxes for the near-disk-center X1.0-class flare (peak: SOL-2024-06-01T18:34, location: S17E24, heliocentric angle: $\theta =$ \ang{29}).
(Top) The observed Fe K$\alpha$ fluorescence flux ($F_{K\alpha}$), measured exciting flux ($F_{>7.11}$), and the theoretically predicted flux for source heights of $h=$ \SI{0}{} and $h=$ \SI{0.1}{R_{\odot}} with photospheric iron abundance of \num{5.5e-5}. The black curve (right axis) shows the plasma temperature evolution.
(Bottom) The temporal fluorescence efficiency ($\Gamma' = F_{K\alpha}/F_{>7.11}$) near the peak of the flare, with the mean value shown by a dotted black line and the uncertainty by the shaded gray region (\num{0.05025 \pm 0.00374}).
The observed Fe K$\alpha$ flux closely tracks both the measured exciting flux and the theoretical predictions, specifically aligning with the simulated flux for a source height ($h=$ \SI{0}{R_{\odot}}).}
    \label{fig:disk_flare_flux}
\end{figure}

\begin{figure}
    \centering
    \includegraphics{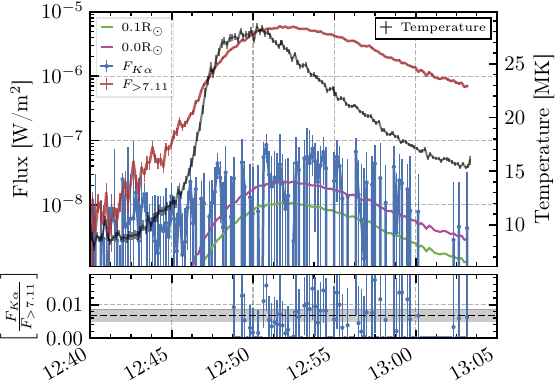}

\caption{Temporal evolution of observed and simulated fluxes for the limb X1.2-class flare (peak: SOL-2024-05-14T12:51, location: S17W89, heliocentric angle: $\theta =$ \ang{89}).
(Top) The observed Fe K$\alpha$ fluorescence flux ($F_{K\alpha}$), measured exciting flux ($F_{>7.11}$), and the theoretically predicted flux for source heights of $h=$ \SI{0}{} and $h=$ \SI{0.1}{R_{\odot}} with photospheric iron abundance of \num{5.5e-5}. The black curve (right axis) shows the plasma temperature evolution.
(Bottom) The temporal fluorescence efficiency ($\Gamma' = F_{K\alpha}/F_{>7.11}$) near the peak of the flare, with the mean value shown by a dotted black line and the uncertainty by the shaded gray region (\num{0.00678 \pm 0.00178}).
In contrast to near-disk-center flare, the observed Fe K$\alpha$ fluorescence flux ($F_{K\alpha}$) remains negligible, despite the presence of comparable exciting flux ($F_{>7.11}$).}

    \label{fig:limb_flare_flux}
\end{figure}

To empirically validate the fluorescence model, the temporal behavior of the derived fluxes is examined throughout the flare evolution. The SoLEXS spectra are integrated into \SI{10}{\second} bins, and each spectrum is fitted with the composite model described in Section~\ref{sect:composite_model}. This procedure yields time series for the Fe K$\alpha$ fluorescence flux ($F_{K\alpha}$), the ionizing flux ($F_{>7.11}$), the plasma temperature ($T$), the emission measure (EM), and the elemental abundances at a \SI{10}{\second} cadence. The time evolution of $T$, $F_{K\alpha}$, and $F_{>7.11}$ for the representative near-disk-center flare are shown in Figure~\ref{fig:disk_flare_flux} (top panel). Both the observed Fe K$\alpha$ flux and the measured exciting flux exhibit a nearly synchronous evolution, with the Fe K$\alpha$ flux peaking at approximately \SI[separate-uncertainty = true]{2.44(0.39)e-7}{\watt \per \metre \squared} in close correspondence with the exciting flux peak of \SI{2.90(0.07)e-6}{\watt \per \metre \squared}.
In contrast, the limb flare (Figure~\ref{fig:limb_flare_flux}) shows negligible Fe K$\alpha$ flux despite having a comparable ionizing flux that peaked near \SI{6.08(0.11)e-6}{\watt \per \metre \squared}. This difference reinforces the geometric dependence of the fluorescence observation.

Furthermore, the observed Fe K$\alpha$ flux is compared with a simulated fluorescence flux derived directly from the theoretical framework of \citet{Bai1979}.
For each \SI{10}{\second} interval, the intrinsic fluorescence efficiency, $\Gamma(T,A_{Fe},h)$, is calculated using the time-resolved best-fit temperatures, fixed photospheric iron abundance of \num{5.5e-5}, and two limiting coronal source heights of $h=$ \SI{0}{} and $h=$ \SI{0.1}{R_{\odot}}
(Section~\ref{sect:fluo_mechanism}).
The expected Fe K$\alpha$ flux for these two heights is then simulated by computing the product of this efficiency, the measured exciting continuum flux ($F_{>7.11}$), and $f(\theta,h)$.
As shown in Figure~\ref{fig:disk_flare_flux}, the observed Fe K$\alpha$ flux closely matches the evolution and magnitude of this simulated flux for a coronal source height of $h=$ \SI{0}{R_{\odot}}.

The observed fluorescence efficiency ($\Gamma'$) is also calculated for each \SI{10}{\second} interval. This quantity, defined as the ratio of the measured Fe K$\alpha$ flux to the total exciting flux ($F_{K\alpha}$/$F_{>7.11}$), accounts for both the intrinsic efficiency of the physical process and the geometric viewing effects ($\Gamma' = \Gamma(T,A_{Fe},h)\cdot f(\theta,h)$).
This observed fluorescence efficiency remains relatively stable for the disk flare (bottom panel of Figure~\ref{fig:disk_flare_flux}). %
Using the inverse-variance weighting method \citep{Bevington2003} across the time series, mean efficiency of \SI{0.05025 \pm 0.00374}{} is calculated during the flare's peak phase.
This stability demonstrates a consistent and direct physical link between the coronal X-ray source and the resulting photospheric fluorescence, allowing for a measurement of the efficiency for each flare.
In contrast, the limb flare exhibits a negligible mean observed fluorescence efficiency of \SI{0.00678 \pm 0.00179}{} (bottom panel of Figure~\ref{fig:limb_flare_flux}), demonstrating the strong suppression of the fluorescence signal at high heliocentric angles.

\subsection{Heliocentric Angle Dependence of Fluorescence Efficiency}
The theoretical framework for iron fluorescence predicts a strong dependence on the viewing geometry. To test this relationship observationally, the observed fluorescence efficiency ($\Gamma'$=$F_{K\alpha}$/$F_{>7.11}$) is calculated for all selected X-class flares and analyzed as a function of heliocentric angle ($\theta$).

\begin{figure}
    \hspace*{.15 \textwidth}
    \includegraphics{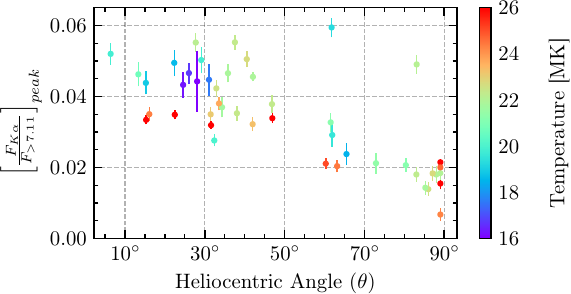}
    \caption{
        Observed fluorescence efficiency
        calculated at the peak of each flare, plotted against its heliocentric angle ($\theta$). Each data point is color-coded by the flare's peak plasma temperature.
        Note that Flare No. 8 (Table~\ref{tab:flare_list}) has been excluded from this plot to better visualize the heliocentric angle dependence, as it exhibits an anomalously high efficiency of \SI{0.0924 \pm 0.0030}{}.
}
    \label{fig:solexs_fluo_all}
\end{figure}

First, the inverse-variance weighted mean of the observed fluorescence efficiency is calculated for each of the \num{47} flares during its peak phase.
The key parameters for each event, including the GOES class, heliographic location, heliocentric angle, peak temperature, and the observed fluorescence efficiency, are tabulated in Table~\ref{tab:flare_list}.
The results are plotted in Figure~\ref{fig:solexs_fluo_all}, with each point color-coded by the flare's peak temperature. The plot shows the expected center-to-limb variation, with the high efficiencies observed for flares near the disk center and a decline for those near the limb. This provides statistical confirmation of the fluorescence model. Additionally, flares with higher peak temperatures (indicated by redder colors) tend to have a comparatively lower fluorescence efficiency. This inverse relationship is consistent with the predictions of the \citet{Bai1979} model (Figure~\ref{fig:fluo_depen}a), which shows that the intrinsic efficiency decreases as the temperature of the coronal source plasma increases. The data exhibit considerable scatter, which is expected as the observed efficiency depends not only on heliocentric angle and temperature but also on the height of the coronal X-ray source, which likely varies from one event to another.

To isolate the geometric dependence from the thermal dependence, a second analysis is performed by selecting data only from time intervals where the coronal source spectrum is similar across different events. The spectral fits are selected where the plasma temperature fell within the range of \SIrange{18.5}{21.5}{\mega\kelvin}. To ensure the analysis captured the gradual thermal phase, only times after the peak of the exciting flux ($F_{>7.11}$) are selected.

\begin{figure}
    \centering
    \includegraphics{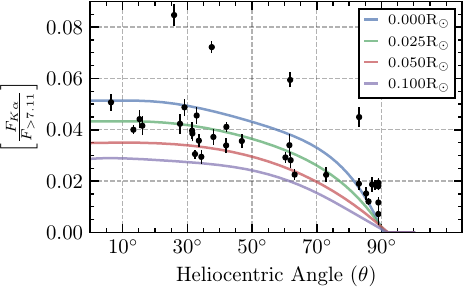}
    \caption{
        Observed fluorescence efficiency
        as a function of heliocentric angle ($\theta$) for the flare sample. To minimize temperature-related effects, each data point is derived from a \SI{10}{second} interval where the flare plasma temperatures are in the range of \SIrange{18.5}{21.5}{\mega\kelvin}. %
        The solid curves represent the theoretically expected efficiency for a \SI{20}{MK} plasma at various coronal source heights (\textit{h}), calculated using the formalism of \citet{Bai1979} with the geometric parameterization from \citet{Drake2008} and a photospheric iron abundance of \SI{5.5e-5}{}.
    }
    \label{fig:solexs_fluo}
\end{figure}

Applying these criteria to the time-resolved (\SI{10}{\second} cadence) spectral fits of the entire sample, \num{34} flares are identified that exhibited sufficient evolution (minimum \SI{100}{\second} or \num{10} samples) within this temperature band during their decay phase. For each of these events, the mean observed fluorescence efficiency is calculated from the qualifying time intervals. Figure~\ref{fig:solexs_fluo} shows these temperature-constrained efficiency values plotted against the heliocentric angle.

The data in Figure~\ref{fig:solexs_fluo} are overlaid with theoretical predictions for a \SI{20}{\mega\kelvin} plasma.
These curves are derived by calculating the intrinsic fluorescence efficiency ($\Gamma$) and the geometric dependence function ($f(\theta,h)$), utilizing the height dependencies modeled by \citet{Drake2008} for several different coronal source heights ($h$).
The absolute magnitude of these theoretical curves is sensitive to the assumed photospheric iron abundance (Figure~\ref{fig:fluo_depen}b). Therefore,
to align the theoretical predictions with the observations and yield positive source heights, a photospheric iron abundance of \SI{5.5e-5}{} is required.
While this is higher than the modern standard photospheric value of \SI{3.16e-5}{} \citep{Asplund2009}, this derived abundance is in exact agreement with the findings of \citet{Parmar1984}, who adopted this value to constrain the heights of flares. Furthermore, the overall magnitude of observed fluorescence efficiencies closely matches the values reported in \cite{Parmar1984}. This analysis confirms that, once the plasma temperature and iron abundance are accounted for, the primary factors governing the observed fluorescence efficiency are the combination of heliocentric angle and coronal source height.

While the majority of events follow the predicted trend, there are four distinct outliers (Flares 8, 17, 19, and 43 in Table~\ref{tab:flare_list}) originating from active regions which produced multiple X-class flares (NOAA 13663, 13664, 13664, and 13848 respectively). These events exhibit anomalously high efficiency and the physical origin of this enhancement is not presently understood and further studies are needed.
Furthermore, flares near the limb ($\theta \approx \ang{90}$) show a systematic excess compared to the theoretical curves. This deviation is likely attributed to the limitations of the theoretical model at grazing angles, specifically the approximation of the photosphere as a geometrically thin but optically thick spherical shell of homogeneous density \citep{Bai1979,Drake2008}.

\begin{table*}[h]
    \centering
    \caption{Parameters of X-class flares observed by SoLEXS. The table lists the date, peak time, GOES class, heliographic location, heliocentric angle ($\theta$), peak plasma temperature ($T_{\text{peak}}$), EM$_{\text{peak}}$, and the observed fluorescence efficiency ($\Gamma' = F_{K\alpha} / F_{>7.11}$) for each event.}
    \label{tab:flare_list}
    \setlength{\tabcolsep}{3.62pt}
    \begin{tabular}{llcccccc}
	\hline
	No. & Time (UT) & Class & Location & $\theta^\circ$ & 
	\vtop{\hbox{$T_{\text{peak}}$}\hbox{(MK)}} & 
	\vtop{\hbox{EM$_{\text{peak}}$}\hbox{($10^{49} \text{cm}^{-3}$)}} & 
	\vtop{\hbox{Observed}\hbox{Efficiency ($\Gamma'$)}} \\

    \hline
    1 & 2024-02-09	13:14 & X3.3 & S37W85 & 86.0 & $23.83 \pm 0.26$ & $18.59 \pm 0.34$ & $0.0139 \pm 0.0019$ \\
    2 & 2024-02-16	06:53 & X2.5 & S16W80 & 80.4 & $24.21 \pm 0.47$ & $14.93 \pm 0.30$ & $0.0207 \pm 0.0019$ \\
    3 & 2024-02-21	23:07 & X1.8 & N17E39 & 42.0 & $25.63 \pm 0.37$ & $8.01 \pm 0.16$ & $0.0322 \pm 0.0020$ \\
    4 & 2024-02-22	06:32 & X1.7 & N16E35 & 38.1 & $24.15 \pm 0.32$ & $7.16 \pm 0.15$ & $0.0353 \pm 0.0022$ \\
    5 & 2024-02-22	22:34 & X6.3 & N17E27 & 31.6 & $37.41 \pm 0.26$ & $28.01 \pm 0.45$ & $0.0319 \pm 0.0011$ \\
    6 & 2024-03-23	01:33 & X1.1 & N27E08 & 28.1 & $16.42 \pm 0.30$ & $7.99 \pm 0.20$ & $0.0442 \pm 0.0085$ \\
    7 & 2024-03-28	20:55 & X1.1 & S14W61 & 61.9 & $20.02 \pm 0.31$ & $6.42 \pm 0.15$ & $0.0292 \pm 0.0033$ \\
    8 & 2024-05-03	02:22 & X1.6 & N25E07 & 25.9 & $27.18 \pm 0.64$ & $6.76 \pm 0.15$ & $0.0924 \pm 0.0030$ \\
    9 & 2024-05-05	05:59 & X1.3 & N26W10 & 27.7 & $24.09 \pm 0.46$ & $6.27 \pm 0.15$ & $0.0552 \pm 0.0026$ \\
    10 & 2024-05-05	11:54 & X1.2 & N26W22 & 33.6 & $24.64 \pm 0.39$ & $5.72 \pm 0.13$ & $0.0380 \pm 0.0019$ \\
    11 & 2024-05-06	06:35 & X4.5 & N25W35 & 42.1 & $25.93 \pm 0.29$ & $23.18 \pm 0.41$ & $0.0455 \pm 0.0013$ \\
    12 & 2024-05-08	01:41 & X1.0 & N26W58 & 61.6 & $23.81 \pm 0.51$ & $4.57 \pm 0.11$ & $0.0327 \pm 0.0027$ \\
    13 & 2024-05-08	05:09 & X1.0 & S18W17 & 24.6 & $17.36 \pm 0.34$ & $6.35 \pm 0.16$ & $0.0433 \pm 0.0036$ \\
    14 & 2024-05-08	21:40 & X1.0 & S20W17 & 26.0 & $17.78 \pm 0.33$ & $6.02 \pm 0.15$ & $0.0466 \pm 0.0030$ \\
    15 & 2024-05-09	09:12 & X2.2 & S20W26 & 32.4 & $20.75 \pm 0.18$ & $12.77 \pm 0.26$ & $0.0276 \pm 0.0017$ \\
    16 & 2024-05-09	17:44 & X1.1 & S14W28 & 31.0 & $18.13 \pm 0.36$ & $6.46 \pm 0.16$ & $0.0447 \pm 0.0046$ \\
    17 & 2024-05-10	06:54 & X3.9 & S17W34 & 37.6 & $25.81 \pm 0.26$ & $22.70 \pm 0.42$ & $0.0552 \pm 0.0021$ \\
    18 & 2024-05-11	01:19 & X5.8 & S15W45 & 46.9 & $30.67 \pm 0.21$ & $29.16 \pm 0.48$ & $0.0339 \pm 0.0014$ \\
    19 & 2024-05-11	11:44 & X1.5 & S19W60 & 61.8 & $20.64 \pm 0.29$ & $9.19 \pm 0.21$ & $0.0594 \pm 0.0028$ \\
    20 & 2024-05-12	16:26 & X1.0 & S18W72 & 72.9 & $22.41 \pm 0.34$ & $5.56 \pm 0.14$ & $0.0212 \pm 0.0029$ \\
    21 & 2024-05-14	02:09 & X1.7 & S17W89 & 89.0 & $30.93 \pm 0.37$ & $7.51 \pm 0.16$ & $0.0155 \pm 0.0016$ \\
    22 & 2024-05-14	12:55 & X1.2 & S17W89 & 89.0 & $27.00 \pm 0.39$ & $6.51 \pm 0.14$ & $0.0068 \pm 0.0018$ \\
    23 & 2024-05-14	16:51 & X8.7 & S18W89 & 89.0 & $44.06 \pm 0.34$ & $32.35 \pm 0.50$ & $0.0215 \pm 0.0010$ \\
    24 & 2024-05-15	14:38 & X2.9 & S12E89 & 89.0 & $23.29 \pm 0.23$ & $17.56 \pm 0.34$ & $0.0184 \pm 0.0019$ \\
    25 & 2024-05-27	07:08 & X2.8 & S18E89 & 89.0 & $25.89 \pm 0.30$ & $13.51 \pm 0.26$ & $0.0201 \pm 0.0016$ \\
    26 & 2024-05-29	14:33 & X1.4 & S19E64 & 65.5 & $19.06 \pm 0.31$ & $8.29 \pm 0.29$ & $0.0238 \pm 0.0031$ \\
    27 & 2024-05-31	22:01 & X1.1 & S17E32 & 35.8 & $27.44 \pm 0.81$ & $5.32 \pm 0.12$ & $0.0465 \pm 0.0026$ \\
    28 & 2024-06-01	08:48 & X1.4 & S18E28 & 32.9 & $24.42 \pm 0.35$ & $6.88 \pm 0.15$ & $0.0423 \pm 0.0025$ \\
    29 & 2024-06-01	18:36 & X1.0 & S17E24 & 29.1 & $22.43 \pm 0.42$ & $4.69 \pm 0.11$ & $0.0503 \pm 0.0037$ \\
    30 & 2024-06-10	11:08 & X1.5 & S19W85 & 85.3 & $23.81 \pm 0.33$ & $7.22 \pm 0.17$ & $0.0143 \pm 0.0020$ \\
    31 & 2024-07-14	02:34 & X1.3 & S12W39 & 40.5 & $25.67 \pm 0.41$ & $5.90 \pm 0.13$ & $0.0505 \pm 0.0023$ \\
    32 & 2024-07-16	13:26 & X1.9 & S05W83 & 83.0 & $24.40 \pm 0.34$ & $10.48 \pm 0.21$ & $0.0181 \pm 0.0021$ \\
    33 & 2024-07-29	02:37 & X1.5 & S05W04 & 6.4 & $24.08 \pm 0.48$ & $6.91 \pm 0.15$ & $0.0520 \pm 0.0030$ \\
    34 & 2024-08-05	13:40 & X1.7 & S10W88 & 88.0 & $22.65 \pm 0.32$ & $7.96 \pm 0.17$ & $0.0180 \pm 0.0024$ \\
    35 & 2024-08-08	19:35 & X1.3 & S04W22 & 22.3 & $19.49 \pm 0.31$ & $7.04 \pm 0.16$ & $0.0495 \pm 0.0037$ \\
    36 & 2024-08-14	06:40 & X1.1 & N13E03 & 13.3 & $21.18 \pm 0.43$ & $5.06 \pm 0.12$ & $0.0462 \pm 0.0034$ \\
    37 & 2024-09-12	09:43 & X1.3 & S16E87 & 87.1 & $25.49 \pm 0.40$ & $5.89 \pm 0.13$ & $0.0184 \pm 0.0021$ \\
    38 & 2024-09-14	15:29 & X4.5 & S16E59 & 60.3 & $25.66 \pm 0.23$ & $22.68 \pm 0.39$ & $0.0211 \pm 0.0015$ \\
    39 & 2024-10-01	22:20 & X7.1 & S16E16 & 22.5 & $31.55 \pm 0.27$ & $32.16 \pm 0.52$ & $0.0349 \pm 0.0013$ \\
    40 & 2024-10-03	12:18 & X9.0 & S15W03 & 15.3 & $33.79 \pm 0.29$ & $35.93 \pm 0.55$ & $0.0334 \pm 0.0013$ \\
    41 & 2024-10-07	19:13 & X2.2 & S16W62 & 63.2 & $26.43 \pm 0.30$ & $9.65 \pm 0.19$ & $0.0204 \pm 0.0017$ \\
    42 & 2024-10-09	01:56 & X1.8 & N13W08 & 15.2 & $19.46 \pm 0.26$ & $10.50 \pm 0.35$ & $0.0438 \pm 0.0032$ \\
    43 & 2024-10-09	15:47 & X1.4 & S10W83 & 83.1 & $24.78 \pm 0.61$ & $5.76 \pm 0.13$ & $0.0490 \pm 0.0026$ \\
    44 & 2024-10-31	21:20 & X2.0 & N15E28 & 31.5 & $24.58 \pm 0.35$ & $8.83 \pm 0.18$ & $0.0350 \pm 0.0021$ \\
    45 & 2024-11-06	13:40 & X2.3 & S08E14 & 16.1 & $28.40 \pm 0.42$ & $9.39 \pm 0.17$ & $0.0350 \pm 0.0019$ \\
    46 & 2024-12-29	07:18 & X1.1 & N10W33 & 34.3 & $24.29 \pm 0.38$ & $5.32 \pm 0.12$ & $0.0370 \pm 0.0028$ \\
    47 & 2024-12-30	04:14 & X1.5 & N10W46 & 46.8 & $25.56 \pm 0.36$ & $7.53 \pm 0.16$ & $0.0378 \pm 0.0025$ \\
    \hline
\end{tabular}
\end{table*}

\section{Discussion and Conclusions} \label{sect:discussion}

The results presented in this study provide a validation of the photospheric iron fluorescence model using the unique capabilities of the Aditya-L1/SoLEXS instrument.
The analysis of \num{47} X-class flares confirms that fluorescence is the dominant production mechanism for the \SI{6.4}{\keV} Fe K$\alpha$ feature during the thermal phase.
This conclusion is supported by two key observations: the temporal correlation between the Fe K$\alpha$ line and the exciting flux (Figure~\ref{fig:disk_flare_flux}), and the center-to-limb variation observed across the flare sample (Figure~\ref{fig:solexs_fluo_all}).

SoLEXS's primary advantage is its ability to measure the exciting flux ($F_{>7.11}$) and the fluorescent line ($F_{K\alpha}$) simultaneously.
Previous studies with narrow-band crystal spectrometers had to rely on model-dependent estimates of the ionizing flux, introducing a significant source of uncertainty. The direct measurement provided by SoLEXS removes this ambiguity, allowing for a more robust calculation of the fluorescence efficiency.
This work, therefore, establishes the capability of modern, broadband SDD spectrometers for quantitative solar fluorescence studies.

While broadband spectrometers like SoLEXS can be insensitive to the continuum emission of cooler plasma hidden beneath a dominant hot flare core \citep{Winebarger2012}, our low temperature contribution analysis (Section~\ref{low_temp_contri}) demonstrates that such a hidden component cannot reproduce the observed line flux at \SI{6.4}{\keV}. The emissivity of the relevant lower-ionization satellite lines (Fe {\sc xix--xxiv}) at \SIrange{6}{10}{\mega\kelvin} is orders of magnitude too low to account for the observed signal without invoking unphysically large emission measures. Furthermore, a two-temperature component analysis  confirms that the contribution from cooler plasma (whether pre-flare background or a secondary flaring component) is negligible, and explicitly modeling this background component does not alter the derived flare parameters or the fluorescence flux.

\begin{figure}[htbp]
    \centering
    \centerline{
        \includegraphics[width=0.47\textwidth,clip=]{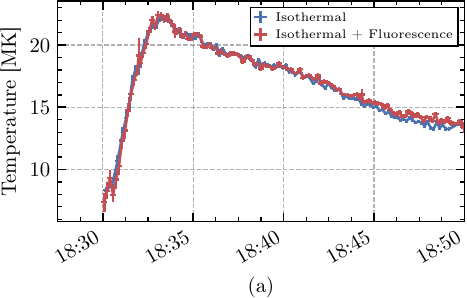}%
        \hspace*{0.05\textwidth}
        \includegraphics[width=0.47\textwidth,clip=]{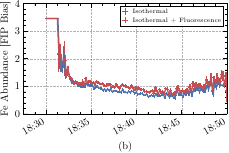}%
    }
    \caption{
        Comparison of best-fit thermal parameters, (a) temperture and (b) iron abundance, derived from the isothermal-only and the composite (isothermal + fluorescence) models.
    }
    \label{fig:thermal_vs_fe}
\end{figure}

Using the representative disk flare of SOL-2024-06-01 (shown in Figure~\ref{fig:disk_flare_analysis_summary}) as a test case, this study also addresses how modeling the fluorescence component affects the derived thermal parameters. As shown in Figure~\ref{fig:thermal_vs_fe}a, the best-fit plasma temperature remains consistent regardless of whether the fluorescence Gaussian components are included in the model. The mean difference between the two fits is only \SI{0.2}{\mega\kelvin}, which is well within the typical fitting uncertainty.
However, a subtle systematic effect is observed for the iron abundance. The thermal-only model tends to slightly underestimate the abundance compared to the composite model (Figure~\ref{fig:thermal_vs_fe}b), with a mean ratio of \num{0.87}. This underestimation is negligible for most purposes. Therefore, for determining the primary thermal properties of the plasma in these X-class flares, a single-component isothermal model provides reliable results, even without explicitly modeling the fluorescence contribution.

By isolating the data to a narrow temperature band, the observed fluorescence efficiency as a function of heliocentric angle shows excellent agreement with the theoretical frameworks established by \citet{Bai1979} and \citet{Drake2008} (Figure~\ref{fig:solexs_fluo}).
    However, to achieve this agreement and yield physically realistic coronal source heights within this analytical framework, a photospheric iron abundance of \SI{5.5e-5}{} was required.
    While this exceeds the modern accepted photospheric value of \SI{3.16e-5}{} \citep{Asplund2009}, it perfectly matches the effective baseline adopted by \citet{Parmar1984} to obtain similar consistency with their observational data.

The high-cadence capabilities of SoLEXS theoretically allow for tracking the temporal evolution of the coronal X-ray source height, a technique previously demonstrated with crystal spectrometers \citep{Parmar1984}.
However, the time-averaged statistical uncertainty in the observed fluorescence efficiency derived in this study is $\approx 0.003$, which is comparable to the uncertainties reported by \citet{Parmar1984}.
Theoretical modeling indicates that this uncertainty corresponds to a source height difference of approximately \num{0.01}{R$_{\odot}$} (for a \SI{20}{\mega\kelvin} point source at disk center, assuming a fixed iron abundance of \num{5.5e-5}).
Consequently, reliably deconvolving height variations on short timescales remains challenging with the current precision, as the uncertainties on individual time bins ($\approx 0.015$) are too large to resolve rapid height changes.

Nevertheless, by utilizing the mean fluorescence efficiency averaged over the flare peak, it is possible to constrain the mean effective coronal source height with a precision of $\approx$ \num{0.01}{R$_{\odot}$}, particularly for disk events.
It is important to note, however,
a fundamental degeneracy exists within this analytical framework that the observed K$\alpha$ flux is simultaneously dependent on both the coronal source height and the photospheric iron abundance, making it impossible to uniquely decouple the two parameters from the fluorescence efficiency alone \citep{Parmar1984}.
Furthermore, this
geometric estimation assumes the fluorescence flux is driven solely by a simple coronal X-ray point source.
While the general trends align with theory, the presence of outliers within the analyzed sample indicates that extended loop geometries, photospheric abundance anomalies, or alternative excitation mechanisms can complicate height determinations. %
Therefore, while this study establishes X-ray fluorescence as the dominant mechanism statistically, deriving precise source heights for individual events requires careful consideration of the specific flare geometry and the inherent height-abundance degeneracy.

Future work will expand this analysis to include the more numerous M- and C-class flares. A larger statistical sample from these lower-intensity events would allow the center-to-limb variation of the fluorescence efficiency to be established with greater precision. Furthermore, while this study confirms fluorescence as the dominant mechanism, some observations have suggested contributions from direct electron impact excitation in certain flares. The continuous, high-duty-cycle observations provided by SoLEXS will produce an unprecedentedly large dataset, offering a unique opportunity to search for and characterize these rarer excitation events.

The precision of this fluorescence efficiency measurement could be significantly enhanced by future instrumentation. While X-ray imaging spectroscopy would be the ideal solution, improvements to non-imaging spectrometers are also feasible. Specifically, increasing the effective area would improve the signal-to-noise ratio of the weak fluorescence line, provided this is matched with the electronics capable of handling the resulting increase in count rates without saturation. Furthermore, improving the spectral resolution to $\approx$ \SI{125}{\electronvolt} and utilizing finer energy binning would allow for a cleaner separation of the \SI{6.4}{\keV} neutral line from the thermal Fe complex, thereby reducing the uncertainties in the fluorescence efficiency measurement.

\begin{acknowledgements}
	The authors express their sincere gratitude to the anonymous referee for their meticulous reading and highly constructive feedback, which have significantly enhanced the clarity and robustness of this paper. A. R. S. extends his deepest thanks to his PhD supervisor, Prof. Santabrata Das, for his invaluable guidance, continuous support, and encouragement.

	Aditya-L1 is an observatory-class mission fully funded and operated by the Indian Space Research Organisation (ISRO). The SoLEXS instrument was designed and developed by the Space Astronomy Group at the U. R. Rao Satellite Centre (URSC). We gratefully acknowledge the extensive collaborative efforts of the various ISRO centers and URSC facilities that realized this payload. Specifically, we thank the URSC teams responsible for thermal systems, mechanisms, systems integration, reliability and quality assurance, and payload assembly. We also extend our gratitude to the Mission and Flight Dynamics Group, ISTRAC, ISSDC, and the Aditya-L1 project team for their roles in mission operations, data management, and dissemination to the scientific community.
\end{acknowledgements}

\bibliographystyle{spr-mp-sola}
\bibliography{references}

\end{document}